\documentclass[aps,pre,superscriptaddress,twocolumn,groupedaddress,showpacs]{revtex4}
\usepackage{graphicx,epsf}
\usepackage{xcolor}
\usepackage{psfrag}
\usepackage{psfrag}

\begin{document}

\title{Loop formation in polymers in crowded environment}
\author{K. Haydukivska}
\affiliation{Institute for Condensed
Matter Physics of the National Academy of Sciences of Ukraine,\\
79011 Lviv, Ukraine}
\author{V. Blavatska}
\email[]{E-mail:  viktoria@icmp.lviv.ua}
\affiliation{Institute for Condensed
Matter Physics of the National Academy of Sciences of Ukraine,\\
79011 Lviv, Ukraine}

\begin{abstract}
We analyze the probability of a single loop formation in a long flexible polymer chain in 
disordered environment in $d$ dimensions. The structural defects are considered to be correlated on 
large distances $r$ according to a power law $\sim r^{-a}$. Working within the frames of 
continuous chain model and applying the direct polymer renormalization scheme, 
we obtain the values of critical exponents governing the scaling of probabilities 
of loop formation with various positions along the chain as function of loops length. 
Our results quantitatively reveal that the presence of structural defects in environment 
decreases the probability of loop formation in polymer macromolecules.
\end{abstract}
\date{\today}

\maketitle

\section{Introduction}

The process when two monomers separated by a large distance along the polymer chain 
come close enough to start interacting with each other is called looping. The loop formation in macromolecules plays an important role in a number of biochemical processes, 
such as stabilization of globular proteins  \cite{Perry84,Wells86,Pace88,Nagi97}, 
transcriptional regularization of genes \cite{Schlief88,Rippe95,Towles09},    
DNA compactification in the nucleus \cite{Fraser06,Simonis06,Dorier09} as well as in number of other 
processes that involve both synthetic and biological polymers.

Statistics of long flexible polymer chains is known to be governed by a set of universal properties, 
independent on any details of microscopic chemical structure of macromolecules \cite{deGennes,desCloiseaux}. In particular, the averaged
 end-to-end distance $\langle R_{N}^2\rangle$ of linear polymer chain obeys the scaling law:
\begin{equation}\label{scalingR}
 \langle R_{N}^2 \rangle \sim N^{2\nu},
\end{equation}
here $\langle \ldots \rangle$ means averaging over an ensemble of possible conformations of macromolecule, $N$ is a number of monomers and $\nu$ is the universal critical exponent.  
In the simplest case of idealized Gaussian chain without any interactions between monomers
$\nu_{{\rm Gauss}}=1/2$, whereas in presence of excluded volume interaction  
this exponent is depending on the space dimension $d$ only:
 $\nu(d{=}2){=}3/4$ \cite{Nienhuis82},
  $\nu(d{=}3){=}0.5882\pm 0.0011$ \cite{Guida98}, $\nu(d{\geq}4){=}1/2$. 
   Moreover, a distance between any two monomers $i$ and $j$ along a polymer chain ($i,j{=}1,\ldots,N$), which are not connected by a chemical bond, scales according to:
\begin{equation}
\langle R_{ij}^2\rangle \sim |i-j|^{2\nu}
\end{equation}
with the same exponent $\nu$.

Let us consider the distribution function $F(R_{ij})$ of the distance $R_{ij}$ between 
any pair of monomers  $i$ and $j$ along a polymer chain. 
In a simplified case of Gaussian polymer,
$F(R_{ij})$ adopts a Gaussian form:
\begin{equation}
F(R_{ij})=\left(\frac{d}{2\pi\langle R_{ij}^2 \rangle}\right)^{d/2}
{\rm e}^{-\frac{d R_{ij}^2}{ 2 \langle R_{ij}^2 \rangle} },
\end{equation}
with
$\langle R_{ij}^2\rangle \sim |i-j|^{2\nu_{{\rm Gauss}}}.
$
The loop in polymer chain corresponds to ``contact'' between two monomers $i$ and $j$, such that $R_{ij}=0$.
For the probability to find a loop of size $|i-j|$  (cyclization probability) in a Gaussian chain we immediately restore the result of Jacobson and Stockmayer \cite{Jacobson50}:
\begin{equation}
P_{{\rm loop}}^{{\rm Gauss}}\equiv F(R_{ij}=0)\sim |i-j|^{-\lambda_{{\rm Gauss}}},
\label{expgaus}
\end{equation}
with $\lambda_{{\rm Gauss}}=-d/2$.

The situation changes drastically when one considers a chain with excluded volume interactions.
The form of $F(R_{ij})$ in this case is much more complicated and depends, in particular, on positions of monomers $i$ and $j$ along the chain \cite{Chan88}. 
The probability to find a loop of size $|i-j|$ in a polymer chain with 
excluded volume interactions scales as
\cite{Redner80}:
\begin{equation}
P_{{\rm loop}}^{ab}\sim |i-j|^{-\lambda_{ab}},\,\,\,\lambda_{ab}=\nu(d+\theta_{ab}) \label{probloop}
\end{equation}
with $a,b=1,2$.
Here, $\lambda_{11}$ refers to the case when both $i$ and $j$ are end monomers, $\lambda_{12}$ -- when $i$ is an end monomer and $j$ is inner one (or vice versa),
$\lambda_{22}$ -- when both $i$ and $j$ are inner monomers (see Fig. 1);
and $\theta_{ab}$ are directly connected with the spectrum of vertex exponents $\sigma$ \cite{Duplantier89}:
\begin{equation}
\theta_{ab}=\sigma_{a}+\sigma_{b}-\sigma_{ab}.
\end{equation} 
In was found, that end monomers of a chain have a higher probability of contact then inner ones \cite{Hsu04,Rubio93,Duplantier89}. 
In particular, in $d{=}3$ the results of numerical simulations  \cite{Hsu04} give: $\theta_{12}{=}0.463$, $ \theta_{22}{=}0.815$, $\theta_{11}{=}0.268.$ and thus:
\begin{eqnarray}
\lambda_{11}=1.920,\,\,\, \lambda_{12}=2.035,\,\,\,\lambda_{22}=2.242.
\end{eqnarray}
The refined analytical studies based on renormalization group approach
give up to the first order of  $\varepsilon{=}4{-}d$ expansion \cite{Duplantier89}:
\begin{eqnarray}
\lambda_{11}=2-\varepsilon/8,\,\,\, \lambda_{12}=2,\,\,\,\lambda_{22}=2+\varepsilon/4. \label{exp}
\end{eqnarray}
Note, that in the case of idealized Gaussian polymer one restores 
the result of Jacobson and Stockmayer (\ref{expgaus}): $\lambda_{11}=\lambda_{12}=\lambda_{22}=\lambda_{{\rm Gauss}}$.

The question of great importance in polymer physics is how the conformational properties of 
macromolecules are modified in presence of structural obstacles (impurities).
One encounters such situations when considering polymers in gels or colloidal 
solutions \cite{Pusey86} or in the crowded environment of biological cells \cite{Kumarrev,Minton01}.
It was shown analytically \cite{Kim83} and confirmed in numerical 
simulations \cite{Kremer,Lee88,Woo91}, that structural disorder in the form 
of randomly distributed point-like defects does not alter the universality class of polymer macromolecules,
unless the the concentration of impurities reaches the percolation threshold   \cite{Kremer,Grassberger93,Ordemann02,Janssen07}. The density fluctuations of defects may 
lead to creation of complex fractal structures \cite{Dullen79}. 
These peculiarities are captured within the model of so-called long-range correlated disorder. Here,
the defects are assumed to be correlated on large distances $r$
according to a power law with a pair correlation function \cite{Weinrib83}:
\begin{equation}
g(r)\sim r^{-a}. 
\end{equation} 
For $a<d$, such a correlation function describes complex (fractal) defects extended in space.
The studies \cite{Blavatska01,Blavatska10,Haydukivska14} quantitatively reveal an extent of the effective size and anisotropy of both linear and closed ring macromolecules  in presence of such a type of disorder.

\begin{figure}[t!]
\begin{center}
\includegraphics[width=80mm]{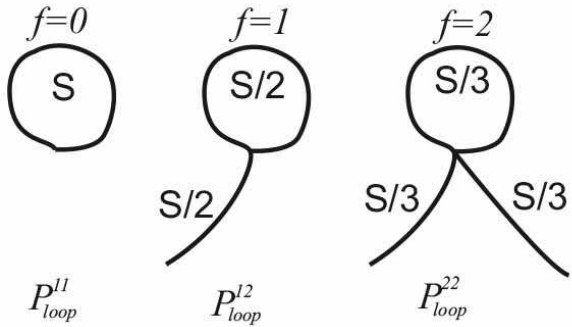}
\caption{ \label{fig:37} Schematic presentation of loops with different positions along the polymer chain. }
\end{center}\end{figure}

 In this concern, it is worthwhile to study a probability of loop formation in the environment with long range correlated disorder,  which has not been considered so far. In the present paper,  we analyze this problem 
analytically  within the frames of continuous chain model, applying the direct polymer renormalization scheme.
The layout of the paper is as follows. In the next Section, we introduce the model. The direct polymer renormalization scheme is shortly described in Section III. We present the results obtained for critical exponents in Section IV and end up with giving conclusions in Section V. 

\section{The Model}

We start with considering a flexible polymer chain in solution in presence of long-range correlated disorder.
Within the Edwards continuous chain model \cite{Edwards}, the chain is presented as a continuous path of length 
$S$, parametrized by $\vec{r}(s)$, where $s$ changes from $0$ to $S$. 
To describe the presence of loop with certain position along the chain  (see Fig. \ref{fig:37}), we consider the system of 
ring polymer connected with $f$ linear chains (with $f=0,1,2$), so that $f=0$ corresponds to a loop formed 
by two end monomers of the chains, $f=1$ -- a loop formed by inner monomer and end one,  
 $f=2$ -- an inner loop. Thus, our problem can be considered as a particular case of 
so-called rosette polymer structures, studied recently in Ref.  \cite{Metzler15}.

 The partition function of the system can thus be presented as:
\begin{eqnarray} 
{\cal Z}_{1;f}(S)=\frac{ {\displaystyle{\int}}\! {\cal {D}}\vec{r}\,\, 
\delta(\vec{r}_1(S)-\vec{r}_1(0)) {\displaystyle {\prod\limits_{i=1}^{1+f_2}}}\delta(\vec{r}_i(0)) 
{\rm e}^{-H}} 
{{\displaystyle {\int}}\! {\cal {D}}\vec{r}\,\,{\displaystyle{\prod\limits_{i=1}^{1+f}}}\delta(\vec{r}_i(0)) 
{\rm e}^{-H}}. \label{model-con} 
\end{eqnarray} 
Here, $\int\! {\cal D}{\vec r}$ denotes functional path integrations over $1+f$ trajectories, the $\delta$-functions describe the fact that one trajectory is closed and that 
the starting point of all $1+f$ trajectories is fixed, 
 and $H$ is a Hamiltonian of the system:
\begin{eqnarray}
&&H=\sum_{i=1}^{1+f}\frac{1}{2}\int^{\frac{S}{1+f}}_{0}{\rm d}s\,\left(\frac{{\rm d}\vec{r}_i(s)}{ds}\right)^{2} +\nonumber\\
&&+\frac{u_0}{2}\sum_{i,j=1}^{1+f}\int^{\frac{S}{1+f}}_{0}{\rm d}s'\int^{s'}_{0}{\rm d}s''\delta(\vec{r}_i(s')-\vec{r}_j(s''))-\nonumber\\
&&-\sum_{i=1}^{1+f}\int^{\frac{S}{1+f}}_{0}{\rm d}s\,V(\vec{r}_i(s)), \label{H-con}
\end{eqnarray}
where the first term describes connectivity of trajectories, 
the second term describes short-range repulsion between monomers due to excluded volume effect governed by coupling constant $u_0$ and the last one
arises due to the presence of disorder in the system and contains a random potential $V(\vec{r}(s))$.
Let us denote by ${\overline{(\ldots)}}$ the average over different realizations of disorder and assume \cite{Weinrib83}:
\begin{equation} {\overline{ V(\vec{r}_i(s'))V(\vec{r}_j(s''))}} = w_0 |\vec{r}_i(s')-\vec{r}_j(s'')|^{-a}, \label{avv0}
\end{equation}
where $w_0$ is a corresponding coupling constant. 
Performing the averaging of the partition function (\ref{model-con}) over different realizations of disorder, taking into account up to the second moment of cumulant expansion and recalling (\ref{avv0}) we obtain
$
{\overline {{\cal Z}_{1;f}(S)}}$
with an effective Hamiltonian:
\begin{eqnarray}
&&H_{eff }=\frac{1}{2}\sum_{i=1}^{1+f}\int_0^{\frac{S}{1+f}}{\rm d}s\!
\left(\frac{{\rm d} {\vec {r}}_i(s)}{{\rm d} s}\right)^2+\\ &&+\frac{u_0}{2}\sum_{i,j=1}^{1+f}\int^{\frac{S}{1+f}}_{0}ds'\int^{s'}_{0}ds''\delta(\vec{r}_i(s')-\vec{r}_j(s''))\,ds-\nonumber\\
 &&-\frac{w_0}{2}\sum_{i,j=1}^{1+f}\int_0^{\frac{S}{1+f}}{\rm d}s'\int_0^{S}{\rm d}s{''}\, |\vec{r}_i(s'')-\vec{r}_j(s')|^{-a}.\nonumber
\label{Hdis}
\end{eqnarray}

The probabilities of loop formations (\ref{probloop}) can be thus calculated as:
\begin{eqnarray}
&&P_{{\rm loop}}^{11}=\frac{{\overline {{\cal Z}_{1;0}(S)}}}{{\overline {{\cal Z}(S)}}}\sim S^{\lambda_{11}},\nonumber\\
&&P_{{\rm loop}}^{12}=\frac{{\overline {{\cal Z}_{1;1}(S)}}}{{\overline {{\cal Z}(S)}}}\sim \left(\frac{S}{2}\right)^{\lambda_{12}},\nonumber\\
&&P_{{\rm loop}}^{22}=\frac{{\overline {{\cal Z}_{1;2}(S)}}}{{\overline {{\cal Z}(S)}}}\sim \left(\frac{S}{3}\right)^{\lambda_{22}},\label{probdef}
\end{eqnarray}
 where ${{\overline {{\cal Z}(S)}}}$ is the total partition function of linear chain of length $S$ given by: 
\begin{eqnarray} 
&&{\overline{{\cal Z}(S)}}=\displaystyle{\int}\! {\cal {D}}\vec{r}\,\exp\left[\frac{1}{2}\int_0^{S}{\rm d}s\!
\left(\frac{{\rm d} {\vec {r}}(s)}{{\rm d} s}\right)^2+\right.\nonumber\\ &&+\frac{u_0}{2}\int^{S}_{0}{\rm d}s'\int^{s'}_{0}{\rm d}s''\delta(\vec{r}(s')-\vec{r}(s''))\,ds-\nonumber\\
&&-\left.\frac{w_0}{2}\int_0^{S}{\rm d}s'\int_0^{S}{\rm d}s{''}\, |\vec{r}(s'')-\vec{r}(s')|^{-a} \right].
 \label{chain} 
\end{eqnarray}

\section{The Method}

We apply the direct renormalization method developed by des Cloizeaux~\cite{desCloiseaux}. 
The main idea of this technique is 
to eliminate various divergences, observed in the asymptotic limit of an infinitely long polymer chain (corresponding to an infinite  number of 
configurations) by introducing corresponding renormalization factors directly associated with physical quantities. 
  In particular, the renormalization factor $\chi_1(\{ z_0 \})$ (with $\{z_0\}$ being the set of bare coupling constants) is introduced via:
 \begin{eqnarray}
\frac{{\cal Z}(S)} { {\cal Z}^{{\rm Gauss}}(S)}=\chi_1(\{ z_0 \}).
   \end{eqnarray}
Here, ${\cal Z}(S)$ is the partition function of single polymer chain and
${\cal Z}^{{\rm Gauss}}(S)$ is the partition function of an idealized Gaussian model.
It is established that the number of all possible conformations of polymer chain scales with the weight of a  macromolecule
parametrised by $S$ as:
\begin{equation}
{\cal Z}(S)\sim \mu^{S}S^{\gamma-1}, \label{gamma1}
\end{equation}
where  $\gamma$ is the universal critical exponent depending only on the space dimension $d$ and  $\mu$ is  a non-universal fugacity.
From the scaling assumption (\ref{gamma1}) one finds
an estimate for an effective critical exponent $\gamma(\{z_0 \})$ governing the scaling behavior of the
 number of possible configurations as:
\begin{equation}
\gamma(\{z_0 \})-1=S\frac{\partial \ln \chi_1(\{z_0 \}) }{\partial S}.\label{gammaexp}
\end{equation}

In a similar way, in our problem we may introduce the factors $\chi_{ab}(\{z_0 \})$ via:
\begin{equation}
\frac{P_{{\rm loop}}^{ab}}{P_{{\rm loop}}^{{\rm Gauss}}}=\chi_{ab}(\{z_0 \})
\end{equation}
with $a,b=1,2$.
Recalling the scaling of loop probabilities 
(\ref{probdef}) and remembering the fact that 
in Gaussian approximation $P_{{\rm loop}}^{{\rm Gauss}}\sim S^{-d/2}$,
one finds an estimate for the effective critical exponents $\lambda_{ab}(\{z_0 \})$:
\begin{equation}
\lambda_{ab}(\{z_0 \})-d/2=-S \frac{\partial \ln \chi_{ab}(\{z_0 \}) }{\partial S}\,.\label{nuexp}
\end{equation}

The critical exponents  presented in the form of series expansions in the coupling constants
are divergent in the asymptotic limit of large $S$.
 To eliminate these divergences, the renormalization of the coupling constants is performed.
The critical exponents attain finite values when evaluated at a stable fixed point  of the renormalization group transformation.
 The flows of the renormalized coupling constants are governed by functions $\beta_{z_{\mathrm{R}}}$:
\begin{equation}
\beta_{z_{\mathrm{R}}}=2S\frac{\partial z_{\mathrm{R}}(\{ z_0\})}{\partial S}\,.
\end{equation}
 Reexpressing $\{ z_0\}$ in terms of renormalized couplings $z_{\mathrm{R}}$,  the fixed points of renormalization group transformations are given
 by common zeros of the $\beta$-functions.
 Stable fixed points govern the asymptotical scaling properties of macromolecules and gives
 reliable asymptotical values of the critical exponents  (\ref{nuexp})	.

\section{Results}

We start with evaluating the partition function ${\cal Z}_{1;f}(S)$  in the simplified Gauusian case, which is then given by:
\begin{eqnarray}
{\cal Z}_{1;f}(S)=\frac{ {\displaystyle{\int}}\! {\cal {D}}\vec{r}\,\, 
\delta(\vec{r}_1(S)-\vec{r}_1(0)) {\displaystyle {\prod\limits_{i=1}^{1+f_2}}}\delta(\vec{r}_i(0)) 
{\rm e}^{ -H_0 }} 
{{\displaystyle {\int}}\! {\cal {D}}\vec{r}\,\,{\displaystyle{\prod\limits_{i=1}^{1+f}}}\delta(\vec{r}_i(0)) 
{\rm e}^{-H_0}}. \label{model-cong} 
\end{eqnarray}
where $ H_0 = \frac{1}{2}\sum\limits_{i=1}^{1+f}\int\limits_0^{S/(1+f)}{\rm d}s\,
\left(\frac{{\rm d} {\vec {r}}_i(s)}{{\rm d} s}\right)^2 $. Using the Fourier-transform of the  $\delta$-functions
\begin{equation}
\delta(\vec{r}_1(S)-\vec{r}_1(0)) =(2\pi)^{-d}\int {\rm d}\vec{q}_1\, {\rm e}^{-i\vec{q}_1(\vec{r}_1(S)-\vec{r}_1(0))}\,,
\end{equation}
we receive 
\begin{equation}
{\cal Z}_{1;f}(S)=(2\pi)^{-d} \int {\rm d}\vec{q}_1\, {\rm e}^{-\frac{q_1^2S}{2}}=(2 \pi S)^{-d/2}.
\end{equation}
In Gaussian approximation the value of partition function does not depend on $f$, and thus the looping probability 
is independent on the loop position along the chain according to Ref. \cite{Jacobson50}.

\begin{figure}[t!]
\begin{center}
\includegraphics[width=60mm]{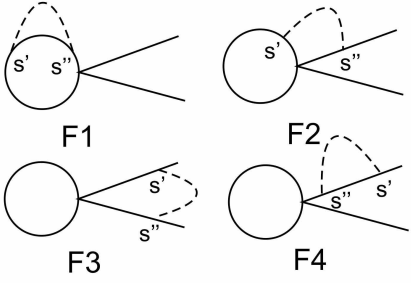}
\caption{ \label{fig:38} Diagrammatic presentation of  contributions into the partition function $Z_{1;2}$ up to the
first order in the coupling constants.}
\end{center}\end{figure}

Taking into account the excluded volume effect and presence of disorder, we present
the partition function as perturbation theory series in dimensionless coupling constants $z_u{=}u_0(2\pi)^{-d/2}S^{2-d/2}$ and $z_w{=}w_0(2\pi)^{-a/2}S^{2-a/2}$:
\begin{equation}
{\cal Z}_{1;f}(S)=(2 \pi S)^{-d/2}-z_uZ^u_{1;f}+z_wZ^w_{1;f}+\ldots
\end{equation}
It is convenient to present the contributions into  $Z^u_{1;f}$ and $Z^w_{1;f}$ using the diagrammatic technique (Fig. \ref{fig:38}). 
Here, dotted lines denote possible interactions between points $s'$ and $s''$ governed by couplings $z_u$ and $z_w$, and integrations are to be performed over all positions of the 
segment end points.  
Note that contributions  into $Z_{1;0}$ are given by only the first diagram and correspond to the partition function of the ring polymer \cite{Haydukivska14}.
On the other hand, contributions into partition function $Z_{1;1}$ are given by all diagrams except $F_3$.
The analytical expressions corresponding to contributions of diagrams $F_1$-$F_4$ into $Z^u_{1;f}$ read:
\begin{eqnarray}
&&{F_1}^u_f=z_u \int_0^{S/(1+f)} {\rm d} s' \int^{s'}_0 {\rm d}s'' (2 \pi)^{-2d} \times\nonumber\\
&&\times\int {\rm d}\vec{q}\, {\rm d} \vec{r}\,\exp \left(-\frac{\vec{q}^2}{2}S_f-\frac{\vec{p}^2}{2}(s'-s'')-\vec{q}\vec{p}(s'-s'')\right),\nonumber\\
&&{F_2}^u_f=z_u \int_0^{S/(1+f)}\, {\rm d}s' \int^{S_f}_0\, {\rm d}s'' (2 \pi)^{-2d} \times\nonumber\\
&&\times\int \,{\rm d}\vec{q}\, {\rm d} \vec{r}\, \exp \left(-\frac{\vec{q}^2}{2}S_f-\frac{\vec{p}^2}{2}(S_f-s''+s')-\vec{q}\vec{p}s'\right),\nonumber\\
&&{F_3}^u_f=z_u \int_0^{S/(1+f)} \,{\rm d}s' \int^{S_f}_0 \,{\rm d}s'' (2 \pi)^{-2d} \times\nonumber\\
&&\times\int \,{\rm d}\vec{q} \,{\rm d} \vec{r}\,\exp \left(-\frac{\vec{q}^2}{2}S_f-\frac{\vec{p}^2}{2}(s'+s'')\right),\nonumber\\
&&{F_4}^u_f=z_u \int_0^{S/(1+f)}\, {\rm d}s' \int^{s'}_0 \,{\rm d}s'' (2 \pi)^{-2d} \times\nonumber\\
&&\times\int \,{\rm d}\vec{q} \, {\rm d} \vec{r}\, \exp \left(-\frac{\vec{q}^2}{2}S_f-\frac{\vec{p}^2}{2}(s'-s'')\right).\nonumber
\end{eqnarray}
Performing the integrations, we get:
\begin{eqnarray}
&&{F_1}^u_f=z_u  \left(2 \pi \frac{S}{1+f}\right)^{-d/2} B\left(1-\frac{d}{2},2-\frac{d}{2}\right),\nonumber\\
&&{F_2}^u_f = z_u  \left(2 \pi \frac{S}{1+f}\right)^{-d/2}\left[\left(1-\frac{d}{2}\right)^{-1} \times \right.\nonumber\\ 
&&\left.\times\left(\frac{5}{4}\right)^{1-d/2}
 H(1/2;-1+d/2,3/2;1/5)- \right.\nonumber\\ 
&&\left.-\left(1-\frac{d}{2}\right)^{-1} B\left(2-\frac{d}{2},2-\frac{d}{2}\right)\right],\nonumber\\
&&{F_3}^u_f=z_u  \left(2 \pi \frac{S}{1+f}\right)^{-d/2} (2^{2-d/2}-2)\times \nonumber\\ 
&&\times\left(1-\frac{d}{2}\right)^{-1}\left(2-\frac{d}{2}\right)^{-1},\nonumber\\
&&{F_4}^u_f=z_u  \left(2 \pi \frac{S}{1+f}\right)^{-d/2} \left(1-\frac{d}{2}\right)^{-1}\left(2-\frac{d}{2}\right)^{-1},\nonumber
\end{eqnarray}
here, $B$ is Euler Beta-function and $H$ is hypergeometric function.

As a result, we obtain contributions into the partition function in the form:
\begin{eqnarray}
&&Z^u_{1;0}=z_u (2 \pi S)^{-d/2} B\left(1-\frac{d}{2},2-\frac{d}{2}\right),\nonumber\\
&&Z^u_{1;1}=z_u  \left(2 \pi \frac{S}{2}\right)^{-d/2}\left[ B\left(1-\frac{d}{2},2-\frac{d}{2}\right)+\right.\nonumber\\
&&\left.+\left(1-\frac{d}{2}\right)^{-1}\left(2-\frac{d}{2}\right)^{-1}-\right.\nonumber\\
&&\left.-\left(1-\frac{d}{2}\right)^{-1}B\left(2-\frac{d}{2},2-\frac{d}{2}\right)+\right.\nonumber\\
&&\left.+\left(1-\frac{d}{2}\right)^{-1} \left(\frac{5}{4}\right)^{1-d/2} H(1/2;-1+d/2,3/2;1/5)\right],\nonumber\\
&&Z^u_{1;2}=z_u  \left(2 \pi \frac{S}{3}\right)^{-d/2}\left[ B\left(1-\frac{d}{2},2-\frac{d}{2}\right)+\right.\nonumber\\
&&\left.+(2^{2-d/2})\left(1-\frac{d}{2}\right)^{-1}\left(2-\frac{d}{2}\right)^{-1}-\right.\nonumber\\
&&\left.-2\left(1-\frac{d}{2}\right)^{-1}B\left(2-\frac{d}{2},2-\frac{d}{2}\right)+\right.\nonumber\\
&&\left. +2\left(1-\frac{d}{2}\right)^{-1} \left(\frac{5}{4}\right)^{1-d/2}H(1/2;-1+d/2,3/2;1/5)\right].\nonumber
\end{eqnarray}
Note, that corresponding contributions into $Z^w_{1;f}$ are obtained simply by substituting $z_u$ by $z_w$ and $d$ by $a$ in expressions in square brackets in above relations.  

Performing a double expansion over parameters  $\varepsilon=4-d$,  $\delta=4-a$ we finally receive the expressions for partition functions for the chains with a loop of certain type:
\begin{eqnarray}
&&\overline{{\cal Z}_{1,0}}(S)=(2 \pi S)^{-d/2} \left(1-z_u\left(\frac{4}{\varepsilon}-2\right)+\right.\nonumber\\
&&\left.+z_w\left(\frac{4}{\delta}-2\right)\right),\nonumber\\
&&\overline{{\cal Z}_{1,1}}(S)=\left(2 \pi \frac{S}{2}\right)^{-d/2} \left(1-z_u\left(\frac{6}{\varepsilon}-1+\right.\right.\nonumber\\
&&\left.+\frac{2}{\sqrt{5}}(\ln2 - \ln(3+\sqrt{5}))\right)+\nonumber\\
&&\left.+z_w\left(\frac{6}{\delta}-1+\frac{2}{\sqrt{5}}(\ln2 - \ln(3+\sqrt{5}))\right)\right),\nonumber\\
&&\overline{{\cal Z}_{1,2}}(S)=\left(2 \pi \frac{S}{3}\right)^{-d/2} \left(1-z_u\left(\frac{10}{\varepsilon}+1-\right.\right.\nonumber\\
&&\left.- \ln 2+\frac{2}{\sqrt{5}}(\ln2 - \ln(3+\sqrt{5}))\right)+z_w\left(\frac{10}{\delta}+\right.\nonumber\\
&&\left.\left.+1-\ln 2+\frac{2}{\sqrt{5}}(\ln2 - \ln(3+\sqrt{5}))\right)\right).
\end{eqnarray}
In addition, the total partition function of linear chain of length $S$ is given by:
\begin{equation}
\overline{{\cal Z}}(S)= 1-z_u\left(-\frac{2}{\varepsilon}-1\right)+z_w\left(-\frac{2}{\delta}-1\right).
\end{equation}
Probabilities of loops formations  are received  using the definition (\ref{probdef}) and read:
\begin{eqnarray}
&&P^{11}_{{\rm loop}}=(2 \pi S)^{-d/2} \left(1-z_u\left(\frac{6}{\varepsilon}-1\right)+\right.\nonumber\\
&&\left.+z_w\left(\frac{6}{\delta}-1\right)\right),\nonumber\\
&&P^{12}_{{\rm loop}}=\left(2 \pi \frac{S}{2}\right)^{-d/2} \left(1-z_u\left(\frac{8}{\varepsilon}+\frac{2}{\sqrt{5}}(\ln2 -\right.\right.\nonumber\\
&&\left.\left.-\ln(3+\sqrt{5}))\right)+z_w\left(\frac{8}{\delta}+\frac{2}{\sqrt{5}}(\ln2 - \right.\right.\nonumber\\ &&\left.\left.-\ln(3+\sqrt{5}))\right)\right),\nonumber\\
&&P^{22}_{{\rm loop}}=\left(2 \pi \frac{S}{3}\right)^{-d/2} \left(1-z_u\left(\frac{12}{\varepsilon}+2-\right.\right.\nonumber\\
&&\left. -\ln 2+\frac{2}{\sqrt{5}}(\ln2 - \ln(3+\sqrt{5}))\right)+\nonumber\\
&&\left.+z_w\left(\frac{12}{\delta}+2- \ln 2+\right.\right.\nonumber\\
&&\left.\left.+\frac{2}{\sqrt{5}}(\ln2 - \ln(3+\sqrt{5}))\right)\right).
\end{eqnarray}

Recalling that (\ref{nuexp}) can be presented in the form
\begin{equation}
\lambda_{ab} - d/2 = -\frac{\varepsilon}{2} \frac{1}{\chi_{ab}} \frac{\partial\chi_{ab}}{\partial z_u}z_u - \frac{\delta}{2} \frac{1}{\chi_{ab}} \frac{\partial\chi_{ab}}{\partial z_w}z_w \label{sigma}
\end{equation}
we find for the critical exponents governing the scaling of looping probabilities:
\begin{eqnarray}
&&\lambda_{11}=(4-\varepsilon/2)+\frac{\varepsilon}{2} \left(\frac{6}{\varepsilon}-1\right) z_u-\frac{\delta}{2} \left(\frac{6}{\delta}-1\right) z_w\nonumber\\
&&\lambda_{12}=(4-\varepsilon/2)+\frac{\varepsilon}{2} \left(\frac{8}{\varepsilon}+\frac{2}{\sqrt{5}}(\ln2 -\right.\nonumber\\ 
&&\left.-\ln(3+\sqrt{5}))\right) z_u-\nonumber\\
&&-\frac{\delta}{2} \left(\frac{8}{\delta}+\frac{2}{\sqrt{5}}(\ln2 - \ln(3+\sqrt{5}))\right) z_w,\nonumber\\
&&\lambda_{22}=(4-\varepsilon/2)+\frac{\varepsilon}{2} \left(\frac{12}{\varepsilon}+2- \ln 2+\right.\nonumber\\
&&\left.+\frac{2}{\sqrt{5}}(\ln2 - \ln(3+\sqrt{5})) \right) z_u-\nonumber\\
&&-\frac{\delta}{2} \left(\frac{12}{\delta}+2- \ln 2+\right.\nonumber\\
&&+\left.\frac{2}{\sqrt{5}}(\ln2 - \ln(3+\sqrt{5}))\right) z_w.
\end{eqnarray}
Taking into account that $z_u$ and $z_w$ in the first order of perturbation theory 
are proportional to $\varepsilon$, $\delta$, and keeping only contributions up to $\varepsilon$ and  $\delta$, the above expressions can be presented as:
\begin{eqnarray}
&&\lambda_{11}=2-\varepsilon/2+3 z_u-3z_w,\label{ll1}\\
&&\lambda_{12}=2-\varepsilon/2+4z_u-4z_w,\\
&&\lambda_{22}=2-\varepsilon/2+6z_u-6 z_w.\label{ll}
\end{eqnarray}

\begin{figure}[t!]
\begin{center}
\includegraphics[width=90mm]{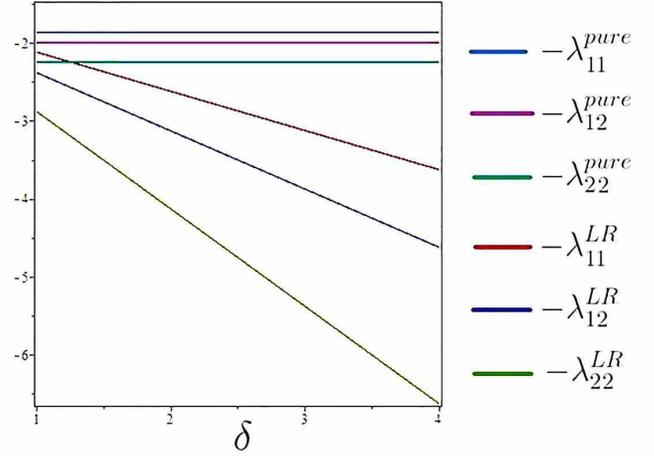}
\caption{ \label{fig:39}Scaling exponents (\ref{purell}) and (\ref{lrll}) as function of $\delta=4-a$ in $d=3$.}
\end{center}
\end{figure}

We make use of results for fixed point values found previously for the linear polymer chains in long-range
correlated disorder \cite{Blavatska01}.  There are three distinct fixed
points governing the properties of macromolecule in various regions of parameters
$d$ and
$a$:\begin{eqnarray}
&& {\rm {Gaussian}}: z^*_{u_0}=0,  z^*_{w_0}=0, \\
&& { \rm {Pure}}: z^*_{u_0}=\frac{\varepsilon}{8}, z^*_{w_0}=0, \\
&& {\rm  {mixed LR}}: z^*_{u_0}=\frac{\delta^2}{4(\varepsilon-\delta)},  z^*_{w_0}=\frac{\delta(\varepsilon-2\delta)}{4(\delta-\varepsilon)}.
\end{eqnarray}
Evaluating Eqs. (\ref{ll1}) - (\ref{ll}) at Gaussian fixed point, we restore the result of 
Jacobson and Stockmayer \cite{Jacobson50}: $\lambda^{{\rm Gauss}}=d/2$.
For polymers with excluded volume effect in  pure solutions we restore results 
of Duplantier \cite{Duplantier89}:
\begin{eqnarray}
&&\lambda^{{\rm pure}}_0=2-\varepsilon/8,\nonumber\\
&&\lambda^{{\rm pure}}_1=2,\nonumber\\
&&\lambda^{{\rm pure}}_2=2+\varepsilon/4.\label{purell}
\end{eqnarray}
Finally,  for the considered case of a polymer chain in disordered environment we received a brand new results:
\begin{eqnarray}
\lambda^{{\rm LR}}_0=2-\varepsilon/2+3\delta/4,\nonumber \\
\lambda^{{\rm LR}}_1=2-\varepsilon/2+\delta,\nonumber \\
\lambda^{{\rm LR}}_2=2-\varepsilon/2+3\delta/2.\label{lrll}
\end{eqnarray}
To find the quantitative estimate for the exponents (\ref{purell}) and (\ref{lrll}) at
$d {=} 3$, we evaluate the expressions at $\varepsilon{=}1$ and various fixed values of $\delta$ (see Fig. \ref{fig:39}). We find,
that  presence of long-range correlated disorder with any
$a {<} d$ leads to an increase of exponents as compared with corresponding pure values, 
and thus the probabilities of loop formation  decreases in presence of crowded environment.
From physical point of view, we
can interpret this as follows. The presence of complex (fractal) obstacles in the system forces the macromolecule to 
avoid these extended regions of space, which results in effective elongation of polymer chain and 
makes the contact of two monomers along the chain less probable.

\section{Conclusions}

In present work we analyzed a probability of loop formation in flexible polymer chains in good solutions, which is known to be governed by scaling laws  (\ref{probloop}) with scaling exponents $\lambda$ dependent on the position of loop along the macromolecule. We considered the special case, when structural obstacles are present in the environment, which are  assumed to be correlated on large distances $r$
according to a power law with a pair correlation function 
$
g(r)\sim r^{-a}$ with  $a<d$ \cite{Weinrib83}. The previous studies  \cite{Blavatska01,Blavatska10,Haydukivska14} reveal 
the non-trivial impact of such a type of disorder on the universal conformational properties of both linear and closed ring macromolecules.

Working within the frames of 
continuous chain model and applying the direct polymer renormalization scheme, 
we obtain the values of critical exponents $\lambda$  up to the first order of perturbation theory in parameters
$\varepsilon=4-d$, $\delta=4-a$.  
Our results quantitatively reveal
that  presence of long-range correlated disorder with any
$a < d$ leads to an increase of exponents as compared with corresponding pure values, 
and thus the probabilities of loop formation  decreases in presence of crowded environment.

\end{document}